\documentstyle[12pt]{article}
\def\be{\begin{equation}}
\def\ee{\end{equation}}
\def\bea{\begin{eqnarray}}
\def\eea{\end{eqnarray}}

\begin{document}

\begin{center}
{\Large{\bf New Form of the T-Duality Due to the Stability
of a Compact Dimension}}

\vskip .5cm
{\large Davoud Kamani}
\vskip .1cm
 {\it Faculty of Physics, Amirkabir University of Technology 
(Tehran Polytechnic)
\\  P.O.Box: 15875-4413, Tehran, Iran}\\
{\sl e-mail: kamani@aut.ac.ir}
\\
\end{center}

\begin{abstract}

We study behaviors of a compact dimension and
the $T$-duality, in the presence of the wrapped closed 
bosonic strings. When
the closed strings interact and form another system
of strings, the radius of compactification increases.
This modifies the $T$-duality,
which we call it as $T$-duality-like. 
Some effects of the $T$-duality-like will be studied.

\end{abstract}

{\it PACS numbers}: 11.25.-w; 11.25.Mj

{\it Keywords}: Compactification; $T$-Duality.

\vskip .5cm
\newpage

\section{Introduction}

Stability of the radius of compactification has been studied from the
various points of view \cite{1}. We look at it from the aspect of
the winding numbers and momentum numbers of the wrapped closed strings.

From the other side, $T$-duality (see, e.g. \cite{2,3})
leads to the ambiguities in the geometry and topology
at the string length scale $l_s =\sqrt{\alpha'}$ \cite{4}.
Consider strings propagating in some background fields (e.g. metric).
These background fields should satisfy the equations of motion. Then,
it turns out that different backgrounds can lead to the same physics without
any observable difference between them. Therefore, this question arises: 
``What is the background metric?'' and hence the background
geometry is ambiguous.

Intuitively, the ambiguities in the geometry arise from the extended nature
of the string. Features in the geometry which are smaller than the 
length scale $l_s$ cannot be detected by using a string
probe whose characteristic size is $l_s$.
The simplest and most widely known example of the ambiguity in the geometry
is the equivalence between a circle with radius $R$ and a circle
with radius $\alpha'/R$. A slightly more peculiar example is the
equivalence between a circle with radius $R=2\sqrt{\alpha'}$ and
a $Z_2$ quotient of a circle (a line segment)
with $R=\sqrt{\alpha'}$. This example demonstrates that the topology
also is ambiguous. Removing these ambiguities leads to the breakdown
of the $T$-duality. Breakdown of the $T$-duality from the nonperturbative
\cite{5} and nonlocality \cite{6} points of view has been studied. 
We also receive it from a different approach.

Some closed strings are winded around a given compact dimension on a 
circle. We assume there are ``$K$'' closed strings on this direction
with the nonzero winding numbers $\{n_1, n_2, \cdot\cdot\cdot, n_K\}$.
These closed strings interact and form another system of closed strings. 
At first, we verify the stability of the compact dimension. 
We find that when at least two strings (before interactions)
have winding numbers with different signs, the compact dimension 
is unstable. Thus, after interactions, the compactification radius
increases. Specially, if all the wrapped closed strings can be released, 
the compact dimension will decompact.

Then, we find the dual radius, $i.e.$ ${\tilde R}$, 
is different from $\alpha'/R$. This modifies the $T$-duality.
We call it as ``$T$-duality-like''.
Consequently, $l_s$ is not minimum length, and the spectrum of
the mass operator on a circle of radius $R$
and the same spectrum on the $T$-dual-like circle, which has
the radius ${\tilde R}$ are different.
In other words, whether a circle respects the usual $T$-duality  
depends on the momentum and winding numbers of the wrapped
closed strings. 

This paper is organized as follows. In section 2, the length of a 
closed string, wrapped around the compact dimension, will be obtained.
In section 3, the stability of the compact dimension, through
the interactions of the initial $K$ closed strings, will be studied.
In section 4, the modification of the $T$-duality for more than
one closed string will be introduced. Section 5 is devoted to 
the conclusions.

Our analysis of a compact dimension depends on the winding
and momentum numbers. These numbers appear in the zero
modes of the bosonic strings.
Thus, we consider only the bosonic strings.
\section{Closed string length}

For a given periodic coordinate $X$ there is the identification
\bea
X \equiv X + 2\pi R ,
\eea
where $R$ is the radius of compactification. A closed string may
now wind around this compact dimension
\bea
X(\sigma + 2\pi , \tau) = X(\sigma , \tau)+2\pi n R , \;\;\; n\in {\bf Z}\;.
\eea
The integer $n$ is the winding number of the closed string.
The momentum of the closed string around this dimension is
quantized
\bea
p=\frac{m}{R} , \;\;\; m \in {\bf Z},
\eea
where the integer ``$m$'' is the momentum number of the closed string.

For the next purposes, we obtain the length of a closed string
with the winding number ``$n$''. The string coordinates 
$X^\mu (\sigma , \tau)$ mean that the string has definite position
and shape. Thus, we can obtain its length.
Let $X^{\bar \mu}$ refers to the non-compact 
space coordinates, $i.e.$
${\bar \mu} \in \{1,2, \cdot\cdot\cdot ,24\}$. The compact coordinate
is $X^{25}=X$. Therefore, the string length is given by
\bea
L_s &~&=\int_{\rm on\;string}\sqrt{(dX)^2 
+ dX^{\bar \mu}dX_{\bar \mu}}
\nonumber\\
&~& =\int_0^\pi d \sigma \sqrt{(\partial_\sigma X)^2 + 
\partial_\sigma X^{\bar \mu}\partial_\sigma X_{\bar \mu}}
\bigg{|}_{\tau=\tau_0}.
\eea
The constant worldsheet time $\tau_0$ indicates 
that the integration is over the string.

A closed string coordinate $X^\mu (\sigma , \tau)$ has the solution
\bea
X^\mu (\sigma , \tau) = x^\mu + 2\alpha' p^\mu \tau 
+2L^\mu \sigma +\frac{i}{2}\ell
\sum_{n \neq 0} \frac{1}{n} \bigg{(} \alpha^\mu_n e^{-2in(\tau-\sigma)}
+{\tilde \alpha}^\mu_n e^{-2in(\tau +\sigma)} \bigg{)},
\eea
where $\ell=\sqrt{2\alpha'}$, and
$L^\mu$ is zero for the non-compact coordinates. For the
compact coordinate $X$, it is $L =nR$. Define $u^\mu (\sigma , \tau_0)$
as in the following
\bea
u^\mu (\sigma , \tau_0)=
\sum_{n \neq 0} \bigg{(} -\alpha^\mu_n e^{-2in(\tau_0-\sigma)}
+{\tilde \alpha}^\mu_n e^{-2in(\tau_0 +\sigma)} \bigg{)}.
\eea
For a closed string with 
nonzero winding number, the length can be written as
\bea
L_s=2L\int_0^\pi d \sigma \bigg{[}1+ \frac{\ell}{L} u +
\bigg{(}\frac{\ell}{2L}\bigg{)}^2 u^2 +
\bigg{(}\frac{\ell}{2L}\bigg{)}^2 u^{\bar \mu}u_{\bar \mu}
\bigg{]}^{1/2}.
\eea
If $\frac{\ell}{L} \ll 1$, then we obtain $L_s \approx n(2\pi R)$.
Neglecting the order $(\frac{\ell}{L})^2$ also leads to the previous
result, $i.e.$,
\bea
L_s = n(2\pi R) + {\cal{O}}(\ell^2/L).
\eea
Therefore, up to the order ${\cal{O}}(\ell^2/L)$,
this is compatible with the equation (2) (e.g. see \cite{7}).
In fact, the equation (2) describes a closed string winding 
$n$ times around the compact direction.
Thus, up to this order, the wrapped 
closed strings completely touch the circle.
In other words, the length of each loop of a wrapped closed string is not
greater than the circle circumference.
\section{Stability of the compactification radius}

Assume there are ``$K$'' wrapped closed strings
with the nonzero winding numbers $\{n_1, n_2, \cdot\cdot\cdot, n_K\}$.
Some of these numbers are positive and the others are negative.
There are also some other closed strings with zero winding number.
Since these strings do not affect the analysis of the system
we shall not consider them.

By the usual splitting-joining process two closed strings with the winding
numbers $n_1$ and $n_2$ can turn into a closed string with the winding
number $N=n_1+n_2$, where each of $n_1$ and $n_2$ may be positive
or negative. In fact, the total winding number is always conserved.

We consider the following processes. $a)$ Strings with winding numbers of
different signs, through the splitting and joining, 
interact and form a set of long strings. Since the un-wrapped strings
do not affect the compactification, we consider only the wrapped closed
strings of these long strings. These long strings
have winding numbers of the same sign. However, the corresponding system
is not the final system.
$b)$ The long and short closed strings in this system interact and form
the final system of closed strings with
the length $2\pi R'$ for each loop of them. In other words,
the final length of each loop determines 
the new radius of compactification $R'$.
When $R' > R$ (or $R'=R$) we say that the {\it initial} compactification
is unstable (stable). After transition to the final system, 
the circumference of the
string loops remains constant. Therefore,
the final radius of compactification $R'$ is stable.

String thermodynamics \cite{8,9} imposes a back-reaction on the
background manifold. This back-reaction gives a dynamics to the spacetime. 
The winding modes of the closed strings
have important rule in this back-reaction to control the decompactification
of the circled dimensions.
That is, the compact dimensions become larger and larger to
touch the wrapped closed strings. Therefore, increasing the radius of 
compactification increases the energy of the wrapped closed strings,
and hence these strings prevent the expansion.

For decompactifying the compact dimension $X$,
all the closed strings with nonzero winding numbers, through the
splitting and joining processes, must interact and produce other closed
strings with zero winding number. This can occur when
the total winding number of the initial system 
\bea
N=\sum_{i=1}^K n_i,
\eea
vanishes. For the case $N \neq 0$ 
the compact dimension is unstable and hence 
the radius of compactification will increases. 

As an example, let the initial system be two closed strings
with the winding numbers $n_1 = +2$ and $n_2 =-1$. 
They transit into a single string
with $N= n_1 +n_2 =+1$. Therefore, the initial radius $R$ changes to
$R'$. The total string length is
$2(2\pi R)+ (2\pi R)$. This length determines the radius $R'$.
It is given by $3(2\pi R)= (2\pi R')$ and hence $R'=3R$.
Similarly, $n_1=-2$ and $n_2=+1$ also lead to $N=-1$, and therefore,
$R'=3R$. For the case $n_1=+2$ and $n_2=+1$, we find $N=+3$.
Thus, we have $3(2\pi R)=3(2\pi R')$ and hence $R'=R$.
In the same way, the case $n_1=-2$ and $n_2=-1$ give $N=-3$ and 
$R'=R$. In the two last cases, the radius is fixed and the winding
number has been increased. In other words, the two last cases
have stable radii.

Now we study the case that $K$ initial closed strings interact
and transit to the final system, $i.e.$, $K'$ (with $1\leq K' \leq N$) 
closed strings with nonzero winding numbers
$\{N_1, N_2, \cdot\cdot\cdot,N_{K'}\}$ and some zero winding strings. 
When the transited system is in its final state
these numbers have the same
sign. Otherwise, more interactions must occur to remove the strings with 
opposite winding numbers.

The total length of the strings in this transition is conserved, $i.e.$,
$|n_1|(2\pi R)+|n_2|(2\pi R)+\cdot\cdot\cdot +|n_K|(2\pi R)=
|N_1(2\pi R')+N_2(2\pi R')+\cdot\cdot\cdot +N_{K'}(2\pi R')|$.
The conservation of the winding numbers also gives 
$\sum_{i=1}^K n_i=\sum_{l=1}^{K'} N_l$. These imply the relation
\bea
|N|R' = R \sum_{i=1}^K |n_i|.
\eea
If $N \rightarrow 0$ the radius $R'$ goes to infinity such that
$|N|R'$ to be finite and nonzero, $i.e.$ equal to $R \sum_{i=1}^K |n_i|$.
However, when $R' \rightarrow \infty$, the dimension $X$ is decompacted.
This occurs when the total winding number $N$ vanishes.
In other words, when $N =0$, the compact dimension
is completely unstable. 
The equation (10) also implies that the ratio $R'/R$ is a rational
number. 

For the next purposes it is useful to write the equation 
(10) in the form
\bea
R' = \frac{\sum_{i=1}^K |n_i|}{|\sum_{i=1}^K n_i|}R.
\eea
Generally, there is the inequality 
$\sum_{i=1}^K |n_i| \geq |\sum_{i=1}^K n_i|$.
Therefore, we have $R' \geq R$. That is, after transition the
radius of compactification increases, or at most remains unchanged.
When all of the initial strings have winding numbers
with the same sign, there is no transition and hence
$R'=R$. In other words, the compact dimension is stable.

The equation (11) is consistent with the argument of \cite{8}.
That is, in the first moment of the Universe, the wrapped strings 
constricted all of the circular dimensions.
These strings were highly likely 
to collide. The collisions involved string/antistring pairs, which
led to annihilations. The collisions continually 
lessened the constriction and
allowed three dimensions to expand.
\section{$T$-duality for $K \geq 2$}

$T$-duality is an intriguing property of string theory. 
In its simplest form, $T$-duality is a remarkable
fact that observable properties (such as the spectrum of
the mass operator) of a closed string compactified on a
circle with the radius $R$ cannot be distinguished from that
of a closed string compactified on a circle with the dual
radius $\alpha'/R$ (where $l_s = \sqrt{\alpha'}$ is the fundamental length
scale of string theory). Effectively, $l_s$ turns out to be
the shortest possible spatial distance, because a shorter
string can always be reinterpreted as a longer string in
the dual theory. 

In the $T$-duality there are the following exchanges
\bea
&~& n \longleftrightarrow m,
\nonumber\\
&~& R \longleftrightarrow \frac{\alpha'}{R}.
\eea
Thus, the left and the right components of the closed string
momentum 
\bea
&~& p_L = \frac{1}{2}\bigg{(}\frac{m}{R}+\frac{nR}{\alpha'}\bigg{)},
\nonumber\\
&~& p_R = \frac{1}{2}\bigg{(}\frac{m}{R}-\frac{nR}{\alpha'}\bigg{)},
\eea
transform to
$p_L \longrightarrow p_L$ and $p_R \longrightarrow -p_R$.
In addition, for the compact dimension, there are the following 
transformations for the oscillators of the closed string coordinate
\bea
&~& {\tilde \alpha}_n \longrightarrow {\tilde \alpha}_n ,
\nonumber\\
&~& \alpha_n \longrightarrow -\alpha_n .
\eea

The closed string mass operator is
\bea
M^2 =\frac{m^2}{R^2} +\frac{n^2 R^2}{\alpha'^2}
-\frac{4}{\alpha'}
+ \frac{2}{\alpha'}\sum^\infty_{n=1}
[(\alpha_{-n}\alpha_n+{\tilde \alpha_{-n}}{\tilde \alpha_n})+
(\alpha^{\bar \mu}_{-n}\alpha_{n{\bar \mu}}+
{\tilde \alpha}^{\bar \mu}_{-n}{\tilde \alpha}_{n{\bar \mu}})].
\eea
Under the transformations (12) and (14)
the mass operator is invariant.
This implies that the geometries with radii $R$ and $\alpha' /R$
are equivalent.
\subsection{Inconsistency of the $T$-duality for $K \geq 2$}

In fact, the T-duality is a symmetry of the string theory
(see sections 2 and 4.1 of the Ref. \cite{2}). This symmetry
is realized by invariance of the partition function of the compactified
string theory under the duality transformations. 
Thus, in the compacted string theory, each equation has a $T$-dual 
version equation with the same feature ($e.g.$ see sections 1 and 4 of 
the Ref. \cite{2}). We demonstrate that for a system with more than one
closed string, $i.e.$ $K \geq 2$, the generality of 
this statement (existence of the T-dual version equation) 
is not true. For example, on the basis of this statement, 
the $T$-dual version of the equation (11) is
\bea
{\tilde {R'}}=\frac{\sum_{i=1}^K |{\tilde n_i}|}{
|\sum_{i=1}^K {\tilde n_i}|} {\tilde R},
\eea 
where ${\tilde R}$ and ${\tilde {R'}}$ 
are $T$-duals of the radii $R$ and $R'$,
respectively. We have ${\tilde n_i}=m_i$ and also ${\tilde m_i}=n_i$.
In addition, if we use ${\tilde R}=\alpha'/R$ and ${\tilde R'}=\alpha'/R'$,
the above equation reduces to
\bea
R' = \frac{|\sum_{i=1}^K m_i|}{\sum_{i=1}^K |m_i|}R.
\eea
This is not correct. To see this, comparing the equations (17) 
and (11) gives
\bea
\frac{\sum_{i=1}^K |n_i|}{|\sum_{i=1}^K n_i|}=
\frac{|\sum_{i=1}^K m_i|}{\sum_{i=1}^K |m_i|}.
\eea
When the integer numbers
$\{n_i\}$ and $\{m_i\}$ are arbitrary, 
this equation does not hold. For example, for
two closed strings with the winding numbers $n_1=+2$ and
$n_2=-1$ this equation gives $3(|m_1|+|m_2|)=|m_1+m_2|$.
For any nonzero numbers $m_1$ and $m_2$ it does not hold.

For solving the problem we should change the $T$-duality transformation of 
the radius $R$. In other words, we shall see that the T-dual version of
$R$, $i.e.$ ${\tilde R}=\alpha'/R$  holds only for the case $K=1$. That is,
for the systems with $K \geq 2$ it is incorrect. 
The correct form of it depends on the
winding numbers and momentum numbers of the closed strings in the initial
system, see the equation (21). 
\subsection{$T$-duality-like transformations}

Since the final radius of the compacted dimension $R'$ is stable,
it obeys the usual $T$-duality exchange 
$R' \longleftrightarrow \alpha'/R'$. 
Due to the initial strings with winding numbers of different signs,
the initial radius $R$ is unstable. Therefore, it does not
obey the usual $T$-duality exchange. According to these, we
introduce the following transformations
\bea
&~& n_i \longleftrightarrow m_i,
\nonumber\\
&~& R' \longleftrightarrow \frac{\alpha'}{R'},
\nonumber\\
&~& R \longrightarrow {\tilde R}.
\eea

Using these transformations and dualizing the equation (11),
lead to the equation
\bea
\frac{\alpha'}{R'} = \frac{\sum_{i=1}^K |m_i|}{|\sum_{i=1}^K m_i|}
{\tilde R}.
\eea
The equations (11) and (20) define the dual radius ${\tilde R}$,
\bea
{\tilde R}=\Omega_K \frac{\alpha'}{R},
\eea
where
\bea
\Omega_K =\frac{|\sum_{i=1}^K m_i||\sum_{i=1}^K n_i|}
{\sum_{i=1}^K |m_i|\sum_{i=1}^K |n_i|}.
\eea 
Therefore, the dual radius of the initial system depends on all momentum
and winding numbers of that system.
Since the numerator of $\Omega_K$ is less 
than or equal to its denominator, we
have $\Omega_K \leq 1$. This gives the inequality
\bea
{\tilde R} \leq \frac{\alpha'}{R}.
\eea 

If the initial system contains closed strings with momentum numbers
of the same sign, and also winding numbers of the same sign,
we obtain $\Omega_K =1$. This system does not have transition,
and hence the compact dimension $X$ is stable.
The dual picture of this system is given by the usual $T$-duality.
If there is one closed string in the initial system, $i.e.$ $K=1$, again
there is no transition. Thus, we have $\Omega_1 =1$
and hence ${\tilde R}=\alpha'/R$, as expected.

In the usual $T$-duality, it is believed that there is no radius
smaller than the self-dual radius ${\tilde R}=R=\sqrt{\alpha'}$.
This indicates the minimum distance scale. In our duality, the
self-dual radius is ${\tilde R}=R=\sqrt{\alpha' \Omega_K}$.
This depends on the momentum and winding numbers of the wrapped
closed strings. Since $\Omega_K \leq 1$, the self-dual radius is
less than $\sqrt{\alpha'}$. In other words, 
detecting the distances less than $\sqrt{\alpha'}$, by two or more
appropriate wrapped closed strings is possible. 
In addition, detecting the distances less than $\sqrt{\alpha'}$,
by one closed string is impossible.

The relation (21) is invariant under the exchanges $n_i \leftrightarrow m_i$
and $R \leftrightarrow {\tilde R}$. Let us write it in the form
$R {\tilde R}=\alpha' \Omega_K$. When $N= \sum_{i=1}^K n_i$
goes to zero, we have $\Omega_K \rightarrow 0$, and hence 
${\tilde R} \rightarrow 0$. To see this, for $N \rightarrow 0$
the equation (11) implies that $R' \rightarrow \infty $.
Therefore, (20) gives ${\tilde R} \rightarrow 0$. Similarly, when 
$\sum_{i=1}^K m_i$ vanishes, the equation (20) leads to
$R' \rightarrow 0$. Thus, from the equation (11) we obtain
$R \rightarrow 0$. However, if the radii ${\tilde R}$ and $R$ are nonzero,
the initial system has nonzero total momentum 
number and total winding number. 

According to (19), 
the left and the right components of the momentum, $i.e.$ (13), transform
as in the following
\bea
&~& p_L \longrightarrow \frac{1}{2}\bigg{(} \frac{1}{\Omega_K}
\frac{nR}{\alpha'} + \Omega_K \frac{m}{R}\bigg{)},
\nonumber\\
&~& p_R \longrightarrow \frac{1}{2}\bigg{(} \frac{1}{\Omega_K}
\frac{nR}{\alpha'} - \Omega_K \frac{m}{R}\bigg{)}.
\eea
These hold for each of the $K$ closed strings.
For simplicity we use $n$ and $m$ instead of $n_i$ and $m_i$.
In terms of the zero modes $\alpha_0 =\ell p_R$ and
${\tilde \alpha}_0 =\ell p_L$, these transformations take the forms
\bea
&~& {\tilde \alpha}_0 \longrightarrow \frac{1}{2}\bigg{[} \bigg{(}
\Omega_K - \frac{1}{\Omega_K}\bigg{)} \alpha_0
+\bigg{(}\Omega_K + \frac{1}{\Omega_K}\bigg{)} {\tilde \alpha}_0
\bigg{]},
\nonumber\\
&~& \alpha_0 \longrightarrow - \frac{1}{2}\bigg{[} \bigg{(}
\Omega_K + \frac{1}{\Omega_K}\bigg{)} \alpha_0
+\bigg{(}\Omega_K - \frac{1}{\Omega_K}\bigg{)} {\tilde \alpha}_0
\bigg{]}.
\eea

Similar to the usual $T$-duality, we generalize these transformations
to the nonzero mode oscillators
\bea
&~& {\tilde \alpha}_n \longrightarrow \frac{1}{2}\bigg{[} \bigg{(}
\Omega_K - \frac{1}{\Omega_K}\bigg{)} \alpha_n
+\bigg{(}\Omega_K + \frac{1}{\Omega_K}\bigg{)} {\tilde \alpha}_n
\bigg{]},
\nonumber\\
&~& \alpha_n \longrightarrow - \frac{1}{2}\bigg{[} \bigg{(}
\Omega_K + \frac{1}{\Omega_K}\bigg{)} \alpha_n
+\bigg{(}\Omega_K - \frac{1}{\Omega_K}\bigg{)} {\tilde \alpha}_n
\bigg{]},
\eea
where $n$ is any integer number.
For $\Omega_K=1$ ($e.g.$ $K=1$), these reduce to the usual
$T$-duality transformations (14), as expected. 
\subsubsection{Transformations of the compact coordinate and the mass
operator}

The compact coordinate of a closed string $X(\sigma , \tau)$ has been
given by (5) without the index $\mu$. According to $p=p_L+p_R$ and
$L= \alpha'(p_L-p_R)$, from the transformations (24) and (26) we obtain
\bea
X \longrightarrow X'&~&=\frac{1}{\Omega_K} x' 
+ \frac{2}{\Omega_K}nR \tau
+2\alpha' \Omega_K \frac{m}{R} \sigma
\nonumber\\
&~& +\frac{i}{4} \ell \sum_{n \neq 0}
\frac{1}{n}\bigg{\{} \bigg{[} \bigg{(}
\Omega_K - \frac{1}{\Omega_K}\bigg{)} \alpha_n
+\bigg{(}\Omega_K + \frac{1}{\Omega_K}\bigg{)} {\tilde \alpha}_n
\bigg{]}e^{-2in(\tau + \sigma)}
\nonumber\\
&~& -\bigg{[} \bigg{(}
\Omega_K + \frac{1}{\Omega_K}\bigg{)} \alpha_n
+\bigg{(}\Omega_K -\frac{1}{\Omega_K}\bigg{)} {\tilde \alpha}_n
\bigg{]}e^{-2in(\tau - \sigma)}\bigg{\}}.
\eea
Since the transformations (24) give 
$p_L +p_R \leftrightarrow \frac{1}{\Omega_K} (p_L - p_R)$,
we also introduced the exchange
\bea
x \longleftrightarrow \frac{1}{\Omega_K} x',
\eea
where $x= x_L + x_R$ and $x'= x_L - x_R$. 
Therefore, the dual coordinate of a string depends on the momentum
and winding numbers of all closed strings, presented
in the system. According to (5), we can
write $X=X_L+X_R$. However, the equation (27) implies 
$X' \neq Y_L-Y_R$ where $Y_L$ and $Y_R$ are functions of
$\tau + \sigma$ and $\tau - \sigma$, respectively. Note that  
(24), (26) and (28) also lead to 
the transformation $X' \rightarrow X$, as expected.

The mass operator (15) under (19) and (26) transforms to
\bea
M^2 \longrightarrow M'^2
&~& =
\bigg{(}\frac{1}{\Omega_K}\frac{n R}{\alpha'}\bigg{)}^2
+\bigg{(}\Omega_K\frac{m}{R}\bigg{)}^2 
-\frac{4}{\alpha'}
+ \frac{1}{\alpha'}\sum^\infty_{n=1} 
\bigg{[}\bigg{(}\Omega_K^2 +\frac{1}{\Omega_K^2}\bigg{)}
(\alpha_{-n}\alpha_n+{\tilde \alpha_{-n}}{\tilde \alpha_n})
\nonumber\\
&~& + \bigg{(}\Omega_K^2 -\frac{1}{\Omega_K^2}\bigg{)}
(\alpha_{-n}{\tilde \alpha}_n+{\tilde \alpha_{-n}}\alpha_n)
\bigg{]}
+\frac{2}{\alpha'}\sum_{n=1}^\infty
(\alpha^{\bar \mu}_{-n}\alpha_{n{\bar \mu}}+
{\tilde \alpha}^{\bar \mu}_{-n}{\tilde \alpha}_{n{\bar \mu}}).
\eea
The mass operator $M^2$ can be written as $M^2 =M^2_L + M^2_R$,
while due to the cross terms, the mass operator $M'^2$ is not
sum of the left and right components.
However, the mass operator $M^2$ (unless  $\Omega_K=1$) is not invariant.
This implies that the geometries with the radii $R$ and ${\tilde R}$
are not equivalent. In other words, geometry is not ambiguous.

For the systems with $\Omega_K = 1$ (e.g. $K=1$), we obtain
$M'^2=M^2$. Therefore, the equations (19), (21) and (24)-(29)
reduce to the usual relations of the usual $T$-duality.
However, for the systems with $\Omega_K \neq 1$, there is 
$M'^2 \neq M^2$ and hence the usual $T$-duality disappears.

For some of the closed string systems our transformations are the
usual $T$-duality transformations. These systems have $\Omega_K=1$.
However, for some other systems these transformations have feature of the 
usual $T$-duality relations, but are different from them. They have
$\Omega_K \neq 1$. Thus, we called the whole possibilities
(i.e. the usual $T$-duality that belongs to $\Omega_K=1$, and
modification of it which has $\Omega_K \neq 1$) as the $T$-duality-like.
\section{Conclusions and summary}

We found that the length of each loop of a closed string, wrapped around
a compact dimension, up to the order $\frac{\ell^2}{L}$ (which is
proportional to $\alpha'/R$), is equal to the circle circumference.
According to the equation (2), this is consistent with the definition
of a compact dimension.
Therefore, changing the length of the closed strings modifies the
radius of compactification. For example $K$ closed strings, 
with positive and negative winding numbers, 
through the processes splitting and joining
produce another system of closed strings. This transition changes
the length of the string loops and hence increases the radius of
compactification. In other words, initial radius is unstable.
The modified radius depends on the winding numbers 
and the initial radius. Thus, the ratio
of the radii is a rational number. If the sum of the initial
winding numbers is zero, the initial compactification completely is unstable.
In this case after transition the compact 
dimension is decompacted. This kind of decompactification process may be 
happened in the first moment of the Universe for some of the dimensions.

We observed that combination of the initial and final 
(transited) systems by the usual
$T$-duality leads to inconsistent relations. The inconsistency 
occurs when the initial system contains more than one closed string.
We suggested a solution as in the following. 
When all closed strings have winding numbers with the same sign,
the compactification radius is stable. Thus, the dual
radius is given by the usual $T$-duality,
$i.e.$ $\alpha'/radius$. For a given closed string  we also considered
the exchange of the momentum number with the winding number, similar to the
usual $T$-duality. When some of the wrapped strings have winding numbers
with different signs, we found that the dual radius
($i.e.$ ${\tilde R}$) of the initial compactification ($i.e.$ $R$)
depends on all momentum and winding numbers of the initial closed strings.
For $K \geq 2$, the dual radius ${\tilde R}$ is less than 
(for special systems, equal to) $\alpha'/R$.

The new form of the dual radius modified the usual duality
transformations of the closed string modes. In other words, under
the $T$-duality-like the compact closed string coordinate does
not transform to $X_L -X_R$. It has a complicated transformation.
In the same way, the closed string mass operator is not invariant.
We conclude that the ambiguities in the geometry and topology disappear.
In addition, we observed that with an appropriate system of closed strings
detecting the distances less than $\sqrt{\alpha'}$ is possible.


\end{document}